\documentclass[a4paper,fleqn,usenatbib]{mnras}
\usepackage{newtxtext,newtxmath}

\usepackage{ae,aecompl}
\usepackage{graphicx}	
\usepackage{amsmath}	
\usepackage{amssymb}	
\usepackage{mathtools}
\usepackage{hyperref}
\usepackage{breqn}
\usepackage{lineno}
\usepackage{relsize}

\title[GAMA: Central/Satellite Quenching]{Galaxy And Mass Assembly (GAMA): Environmental Quenching of Centrals and Satellites in Groups}
\author[L. J. M. Davies et. al.]{L. J. M. Davies\thanks{E-mail:
 luke.j.davies@uwa.edu.au},$^{1}$  A. S. G. Robotham,$^{1,2}$ C. del P. Lagos,$^{1,2}$ S. P. Driver,$^{1}$ \newauthor A. R. H. Stevens,$^{1}$  Y. M. Bah\'{e},$^{3}$ M. Alpaslan,$^{4}$ M. N. Bremer,$^{5}$ M. J. I. Brown,$^{6}$  \newauthor  S. Brough,$^{7}$ J. Bland-Hawthorn,$^{8}$  L. Cortese,$^{1,2}$ P. Elahi,$^{1}$ M. W. Grootes,$^{9}$  \newauthor  B. W. Holwerda,$^{10}$ A. D. Ludlow,$^{1}$ S. McGee$,^{11}$   M. Owers,$^{12}$ S. Phillipps$^{5}$ \\
\\ 
$^{1}$ ICRAR, The University of Western Australia, 35 Stirling Highway, Crawley, WA 6009, Australia\\
$^{2}$ ARC Centre of Excellence for All Sky Astrophysics in 3 Dimensions (ASTRO 3D)\\
$^{3}$ Leiden Observatory, Leiden University, P.O. Box 9513, NL-2300 RA Leiden, the Netherlands \\
$^{4}$ Center for Cosmology and Particle Physics, Department of Physics, New York University, 726 Broadway, New York, NY 10003, USA\\
$^{5}$ School of Physics, University of Bristol, Tyndall Avenue, Bristol BS8 1TL, UK\\
$^{6}$ School of Physics and Astronomy, Monash University, Clayton, Victoria 3800, Australia\\
$^{7}$ School of Physics, University of New South Wales, NSW 2052, Australia\\ 
$^{8}$ Sydney Institute for Astronomy, School of Physics A28, University of Sydney, NSW 2006, Australia\\
$^{9}$ Netherlands eScience Center, Science Park 140, 1098 XG Amsterdam, The Netherlands\\
$^{10}$ Department of Physics and Astronomy, 102 Natural Science Building, University of Louisville, Louisville KY 40292, USA\\
$^{11}$ School of Physics and Astronomy, University of Birmingham, Edgbaston, Birmingham, B15 2TT UK \\
$^{12}$ Department of Physics and Astronomy, Macquarie University, North Ryde, NSW 2109, Australia
\vspace{-5mm}
}

\date{Accepted XXX. Received YYY; in original form ZZZ}

\pubyear{2018}

\begin{document}
\label{firstpage}
\pagerange{\pageref{firstpage}--\pageref{lastpage}}
\maketitle

\begin{abstract}

Recently a number of studies have found a similarity between the passive fraction of central and satellite galaxies when controlled for both stellar and halo mass. These results suggest that the quenching processes that affect galaxies are largely agnostic to central/satellite status, which contradicts the traditional picture of increased satellite quenching via environmental processes such as stripping, strangulation and starvation. Here we explore this further using the Galaxy And Mass Assembly (GAMA) survey which extends to $\sim2$\,dex lower in stellar mass than SDSS, is more complete for closely-separated galaxies ($\gtrsim$95\% compared to $\gtrsim$70\%), and identifies lower-halo-mass groups outside of the very local Universe (M$_{\mathrm{halo}}\sim10^{12}$\,M$_{\odot}$ at $0.1<z<0.2$). As far as possible we aim to replicate the selections, completeness corrections and central/satellite division of one of the previous studies but find clear differences between passive fractions of centrals and satellites. We also find that our passive fractions increase with both halo-to-satellite mass ratio and central-to-second rank mass ratio. This suggests that quenching is more efficient in satellites that are low-mass for their halo ($i.e$ at high halo-to-satellite mass ratio in comparison to low halo-to-satellite mass ratio) and are more likely to be passive in older groups - forming a consistent picture of environmental quenching of satellites.  We then discuss potential explanations for the previously observed similarity, such as dependence on the group-finding method.

\end{abstract}


\section{Introduction}

Galaxies in the local Universe can be broadly classified into two populations: blue star-forming systems, which are forming new stars, and red quiescent (or passive) systems, which have little or no active star-formation \citep[$e.g.$][]{Blanton03, Kauffmann03a, Kauffmann04, Baldry04, Balogh04, Brinchmann04, Taylor15}. Our understanding of galaxy evolution processes suggests that galaxies initially form and then subsequently grow in stellar mass via star-formation and mergers, starting as as blue star-forming systems and then evolving into a quiescent state \citep[$e.g.$][]{Bell04, Faber07, Martin07}. When selected either in rest-frame colour \cite[$e.g.$][]{Baldry04, Taylor15}, or in the specific star-formation rate (sSFR) or star-formation rate (SFR) vs stellar mass (M$_{*}$) plane \cite[$e.g.$][]{Balogh04, Moustakas13, Davies16b, Davies18b}, these populations show clear bimodality; highlighting that the transition from star-forming to quiescent (quenching) is potentially fast \citep[however $c.f.$][]{Schawinski14, Bremer18} and occurs over a broad range of stellar masses. Determining the processes that drive this change is essential to our understanding of galaxy evolution processes. 

Current observational evidence suggests that there are two dominant modes of galaxy quenching. Firstly, secular quenching, which can occur in all galaxies irrespective of external processes and is correlated with the internal properties of a galaxy \citep[][]{Kauffmann03b, Driver06,Wake12,Lang14, Barro17}. This mode of quenching appears to be more pronounced at higher stellar masses (log$_{10}$[M$_{*}$/M$_{\odot}$]$>$10.0), with quenched fraction correlating with the presence of a massive bulge \citep{Fang13, Bluck14, Bremer18}, high central velocity dispersion \citep{Wake12, Teimoorinia16} and/or an Active Galactic Nucleus \citep[AGN, $e.g.$][]{Nandra07}. However, simulations also require gas outflows which inhibit star-formation in the lowest mass galaxies ($e.g.$ log$_{10}$[M$_{*}$/M$_{\odot}$]$\lesssim$9.0) in order to reproduce the observed distribution of low-mass systems \citep{Dekel86}. These outflows are generally attributed to stellar feedback processes \citep{DallaVecchia08}, which at log$_{10}$[M$_{*}$/M$_{\odot}$]$\gtrsim$9.0 do not drive the gas with enough energy to escape the galaxy's gravitational potential  \citep[$e.g.$][]{Dekel86}. This potentially indicates that secular quenching is bimodal with stellar mass - affecting both low and high-mass galaxies but leaving intermediate mass systems relatively unscathed \citep[see][and similar results in the EAGLE simulation from Katsianis et al in preparation]{Davies18b}). 

The second mode of quenching is driven by a galaxy's local environment. Over-dense environments such as clusters, groups or even close pairs \citep[see][]{Patton11, Robotham14, Davies16a} can either remove or inhibit the supply of gas required for ongoing star formation, leading to a quenching event \citep[$e.g.$][]{Peng10, Schaefer17}. There are various physical processes which drive this quenching such as starvation/strangulation \citep{Larson1980, Moore99, Peng15, Nichols11}, tidal and ram-pressure stripping \citep{Gunn72, Moore99, Poggianti17,Brown17, Barsanti18}, and/or harassment \citep{Moore96}. This mode is more likely to affect intermediate-to-low-mass galaxies \citep[log$_{10}$(M$_{*}$/M$_{\odot}$)$<$10.0, $e.g.$ see][]{Davies18b}, and is found to correlate with local galaxy density within groups/clusters \citep{Peng12, Treyer18}, and group/cluster-centric position \citep{Wolf09, Wetzel12, Woo15, Barsanti18} - likely due to the fact that low-mass galaxies moving through over-dense environments cannot retain or accrete star-forming gas. 

However, these environmental processes should only quench satellites and will not generally affect central galaxies that sit at the centre of their haloes (thus are not subject to stripping) and are typically the most massive galaxy in their group (hence tidal interactions /harassment will be minimal, and they can retain their gas). As such, we may expect centrals and satellites to undergo different quenching mechanisms and display different passive fractions when controlled for all other effects \citep[$e.g.$][]{vandenBosch08, Weinmann09, Wetzel12, Peng12, Knobel13,Robotham14, Grootes17}. This is the typically accepted model for environmental quenching processes, where satellite galaxies undergo additional quenching in over-dense environments \citep[$e.g.$][]{Wetzel13, Treyer18}, especially when a satellite is significantly less massive than it's central/halo. This model is both used in numerous galaxy evolution models \citep[$e.g.$][]{Cole00, Henriques15, Stevens17, Cora18, Lagos18b} and observed in hydrodynamic simulations \citep[$e.g.$][]{Bahe15}.

In contrast to this, a number of recent studies have suggested that centrals and satellites show similar passive fractions when controlled for stellar and halo mass, and thus may undergo similar quenching irrespective of their current central/satellite status \cite[$e.g.$][]{Hirschmann14,Knobel15}. Most recently, \cite{Wang18a}, hereafter W18, use SDSS galaxies and the group catalogues of \cite{Yang07}, hereafter Y07,  to explore the passive fraction of central and satellite galaxies as a function of halo mass when controlled for stellar mass and vice-versa. Once controlled, they find that there is no significant difference between the passive fraction of centrals and satellites, and suggest that preferential environmental quenching of satellites at a given stellar and halo mass is not a dominant mechanism in the formation of passive systems. This contradicts the currently held view of satellite quenching. 

However, W18 do find that their similarity between centrals and satellites is most apparent in high-mass galaxies (where environmental quenching is likely to be smallest), and also that their choice of the Y07 group finder may bias their results. For example, \cite{Campbell15} show one of the main tendencies of many group finders is to provide poor designation of centrals/satellites leading to increased artificial central-satellite similarity. In a follow-up paper, \cite{Wang18b} explore the passive fractions of centrals and satellites in the L-GALAXIES semi analytic model \citep{Henriques15} and EAGLE simulations \citep{Schaye15} and find weak to no intrinsic similarity. However, when applying the Y07 group finder to simulated light cones in both simulations, they observe consistent passive fractions between the central/satellite populations. In addition, it is also interesting to note that both \cite{Hirschmann14} and \cite{Knobel15}, who find similar results to W18, also use SDSS galaxies and the Y07 group finding method. This may tentatively suggests that any similarity is driven by the group finding process and is not a true physical effect. Clearly, this warrants further study. 

In this work we perform a direct comparison to the W18 analysis (aiming to replicate their selection and techniques), in order to compare our results to these previous studies. While \cite{Hirschmann14} and \cite{Knobel15} use varying analysis techniques, they essentially use the same data and same group finding method as W18, and arrive at the same results. Therefore, our direct comparison to W18 serves as a comparison to these pervious studies.            

Here we use the Galaxy and Mass Assembly (GAMA) sample to explore the passive fractions of central and satellite galaxies when controlled for stellar mass and halo mass in a similar manner to W18. GAMA extends to $\sim2$\,dex lower in stellar mass than SDSS and is more complete to closely separated galaxies \citep[$>$95\% compared to $>$70\%, see ][]{Liske15}. This allows the identification and parameterisation of both lower mass groups and their satellite populations. Importantly GAMA also uses a completely different method for group finding than Y07, following the friends-of-friends method outlined in \cite{Robotham11}, hereafter, R11. There is one key difference between these group finding methods. The Y07 group finder assigns halo masses to isolated centrals based on abundance matching, while the R11 group finding requires that a halo has at least two members. This leads to the SDSS Y07 group catalogue being dominated by isolated centrals, while the GAMA R11 catalogue is largely composed of satellites. This important distinction may have strong implications for the observed central/satellite passive fractions at fixed stellar and halo mass.  

Throughout this paper we use a standard $\Lambda$CDM cosmology with {H}$_{0}$\,=\,70\,kms$^{-1}$\,Mpc$^{-1}$, $\Omega_{\Lambda}$\,=\,0.7 and $\Omega_{M}$\,=\,0.3.                      
            
\section{Data and Sample Selection}

The GAMA survey second data release (GAMA II) covers 286\,deg$^{2}$ to a main survey limit of $r_{\mathrm{AB}}<19.8$\,mag in three equatorial (G09, G12 and G15) and two southern (G02 and G23 - survey limit of $i_{\mathrm{AB}}<19.2$\,mag in G23) regions. The spectroscopic survey was undertaken using the AAOmega fibre-fed spectrograph \citep[][]{Saunders04,Sharp06} in conjunction with the Two-degree Field \citep[2dF,][]{Lewis02} positioner on the Anglo-Australian Telescope, and obtained redshifts for $\sim$240,000 targets covering $0<z\lesssim0.5$ with a median redshift of $z\sim0.2$, and highly uniform spatial completeness \citep[see][for a summary of GAMA observations]{Baldry10,Robotham10,Driver11}. Full details of the GAMA survey can be found in \citet{Driver11, Driver16},  \citet{Liske15} and \citet{Baldry18}. In this work we utilise the data obtained in the 3 equatorial regions, which we refer to here as GAMA II$_{Eq}$.

In this work we limit our sample to galaxies that have a confirmed, local-flow-corrected redshift at $0.01<z<0.2$ and are not classified as an AGN using the Baldwin, Phillips \& Terlevich diagnostic \citep[BPT,][]{Bladwin81} and the starforming-AGN dividing line of \cite{Kauffmann03c}, as AGN contribution to H$\alpha$ lines can significantly bias SFR measurements. For details of this process for the GAMA sample, see \cite{Davies15b}. However, note that here we also repeat our analysis without excluding AGN sources and our results do no change.    

\subsection{Stellar Masses and SFRs}

Stellar masses for the GAMA II$_{Eq}$ sample are derived from the $ugriZYJHK$ photometry using a method similar to that outlined in \cite{Taylor11} - assuming a Chabrier IMF \citep{Chabrier03}. The SFRs used in this work are presented in \cite{Davies16b}. We primarily use H$\alpha$-derived SFRs to be consistent with W18 \citep[however, note the differences between SDSS and GAMA emission line measurements due to different fibre size, see][]{Hopkins13}. These are measured using GAMA spectra discussed in \citet{Liske15} and the process outlined in \cite{Gunawardhana11,Gunawardhana15} and \cite{Hopkins13}, and using the line measurements of \cite{Gordon17}. However, in Appendix \ref{sec:variation} we also reproduce our analysis using the \textsc{magphys}-derived SFRs outlined in \cite{Driver18}, which are based on the energy balance Spectral Energy Distribution (SED)-fitting code \textsc{magphys} \citep{daCunha08}. Full details of these SFR indicators are described at length in \cite{Davies16b}. All photometry used for these stellar masses and SFRs are measured using the Lambda Adaptive Multi-Band Deblending Algorithm for R (\textsc{lambdar}) and presented in \cite{Wright16}.

\subsection{Group Centrals And Satellites}
\label{sec:Groups}

For our halo masses we use the G$^{3}$C catalogue which includes the identification of all galaxy groups and pairs within GAMA \citep[][see also \citealp{Robotham12,Robotham13,Robotham14, Davies15b}]{Robotham11}. Briefly, the GAMA group catalogue is produced using a bespoke friends-of-friends based grouping algorithm, which was tested extensively on mock GAMA galaxy light cones, and assigns $\sim40\%$ of GAMA galaxies to multiplicity N$>$1 pairs and groups. In this work we define a group as a system with multiplicity N$>$1 ($i.e.$ we include both pairs and groups; we do also repeat our analysis excluding pairs, but find no significant difference in our results). In Appendix \ref{sec:singles} we also discuss how including $N=1$ systems (isolated centrals) affects our results and provides a potentially more direct comparison to the results using the Y07 finder.  

Group halo masses are calculated by group matching to bespoke simulated light cones using a number of methods. Here we use the scaled mass proxy, $M_{\mathrm{halo}}\sim AR_{50}\sigma^2$. Where $R_{50}$ is the radius containing 50\% of the group members, $\sigma$ is the group velocity dispersion and $A$ is a functional scaling factor based on group multiplicity and redshift (see Section 4.3 of R11). For full details of the group finding and mass estimates, see R11. Note that we have also repeated our analysis using the weak-lensing recalibrated group halo masses outlined in \cite{Viola15}. While our individual passive fractions controlled for stellar and halo mass do change when using these halo masses (specifically at low halo masses), we do still see a clear separation between centrals and satellites, and all of the trends in our results remain.    

Critical to the analysis of differences between central and satellite galaxies is the choice of group central. The G$^{3}$C catalogue gives a number of different group centrals based on various approaches. Firstly, the catalogue provides a central estimate based on the source with the brightest $r$-band luminosity of the group members (the brightest cluster galaxy, or BCG), this is comparable to the group central from Y07 and used in W18, \cite{Hirschmann14} and \cite{Knobel15}. For the main analysis in this paper this is the definition of central we use. However, in Appendix \ref{sec:variation} we also discuss how our results vary with choice of central. For these we use both the galaxy closest to the $r$-band centre of light and the galaxy closest to the iterative centre of the group. The iterative centre is defined where, in iterative steps, the $r$-band centre of light is calculated and the most distant galaxy rejected until only two galaxies remain. Then the brightest galaxy is selected as the central. In practice the BCG and iterative central provide similar results, while the centre of light deviates significantly; specifically within high multiplicity groups. See R11 for a full description of these central definitions.      

\begin{figure}
\begin{center}
\includegraphics[scale=0.5]{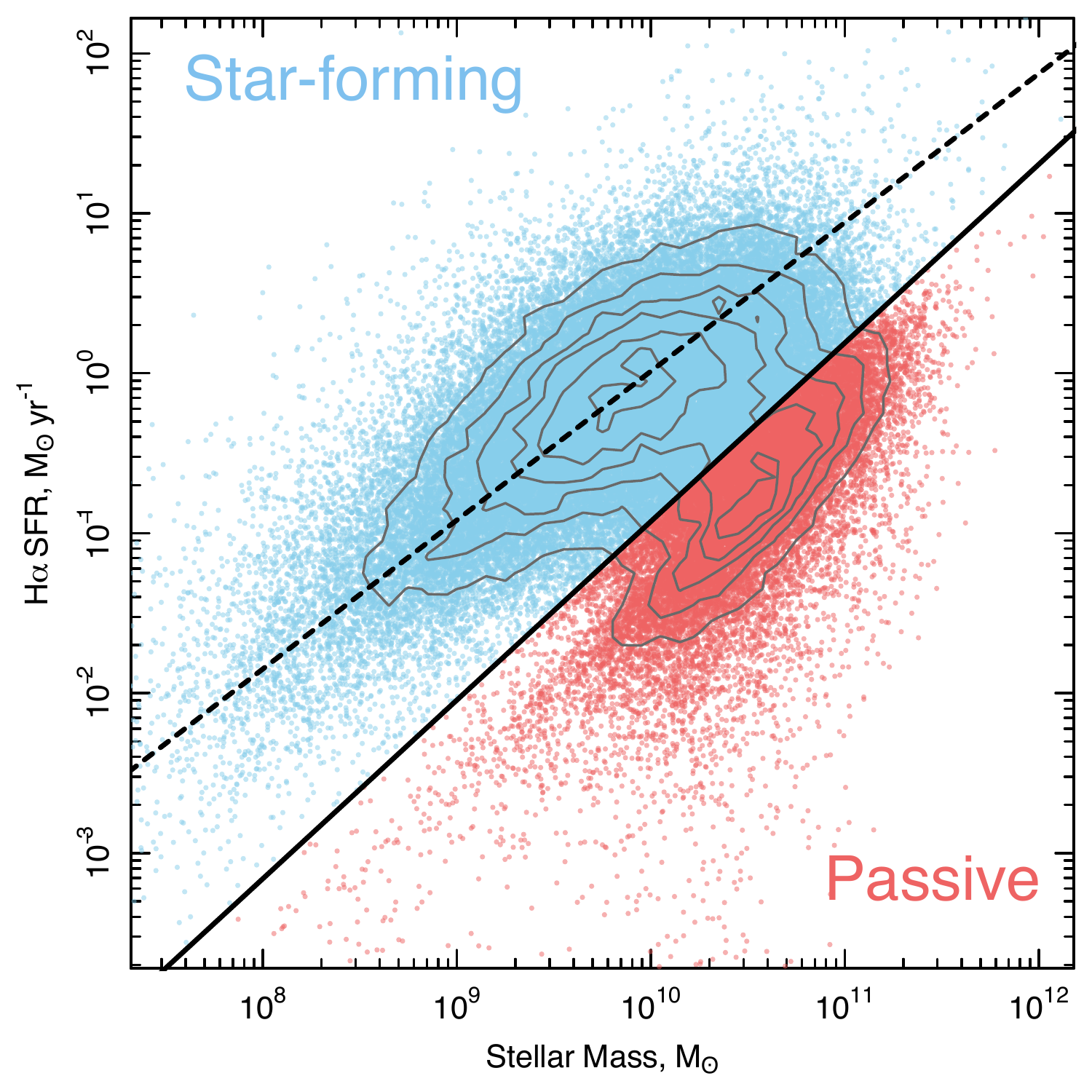}
\vspace{-4mm}

\caption{Selection of star-forming and passive galaxies using the SFR-M$^{*}$ plane. The dashed line displays a fit to the star-forming sequence, while the solid line displays our dividing line between the two populations. Star-forming galaxies are shown in blue and passive galaxies in red. Contours display the density of points in the full GAMA $z<0.2$ non-AGN sample. }
\label{fig:selection1}
\end{center}
\end{figure}

\subsection{Selecting Passive/Star-forming Galaxies}

Multiple methods are available for distinguishing between passive and star-forming systems based on SFR, morphology, structure, etc. \citep[$e.g.$][]{Davies18b}. These can also have a significant impact on derived results depending on the exact selection used and the method for measuring SFRs (for example H$\alpha$-derived SFRs vs SED-derived SFRs). Here we wish to be consistent with W18 who use H$\alpha$-derived SFRs from the New York University Value Added Galaxy Catalogue \citep[NYU-VAGC,][]{Blanton05} and separate star-forming and passive systems using an offset from the star-forming sequence (SFS). 

Fig. \ref{fig:selection1} displays the SFR-M$_{*}$ plane for all $0.01<z<0.2$ non-AGN GAMA galaxies. First we exclude sources with sSFR$<$10$^{-10.5}$\,yr$^{-1}$ and derive a least squares regression fit to the SFS (dashed line). W18 opt to divide passive and star-forming systems at 1\,dex below the SFS. However, within GAMA we find that the passive cloud does not have the same slope as the SFS. Hence, for a more robust dividing line we also derive a least squares regression fit to passive sources with sSFR$<$10$^{-10.5}$\,yr$^{-1}$. To define the dividing line between the two populations, we first take the minimum density point in cross-sections along lines of the shortest distance between our star-forming and passive fits. We then fit these minimum points and take this as our dividing line (black solid line in Fig. \ref{fig:selection1}). The contours in Fig. \ref{fig:selection1} show the density of points for all GAMA galaxies and highlight that the black solid line traces the minimum ridge. The effect of changing our choice of this dividing line is described in Appendix \ref{sec:variation}.

\begin{figure}
\begin{center}
\includegraphics[scale=0.55]{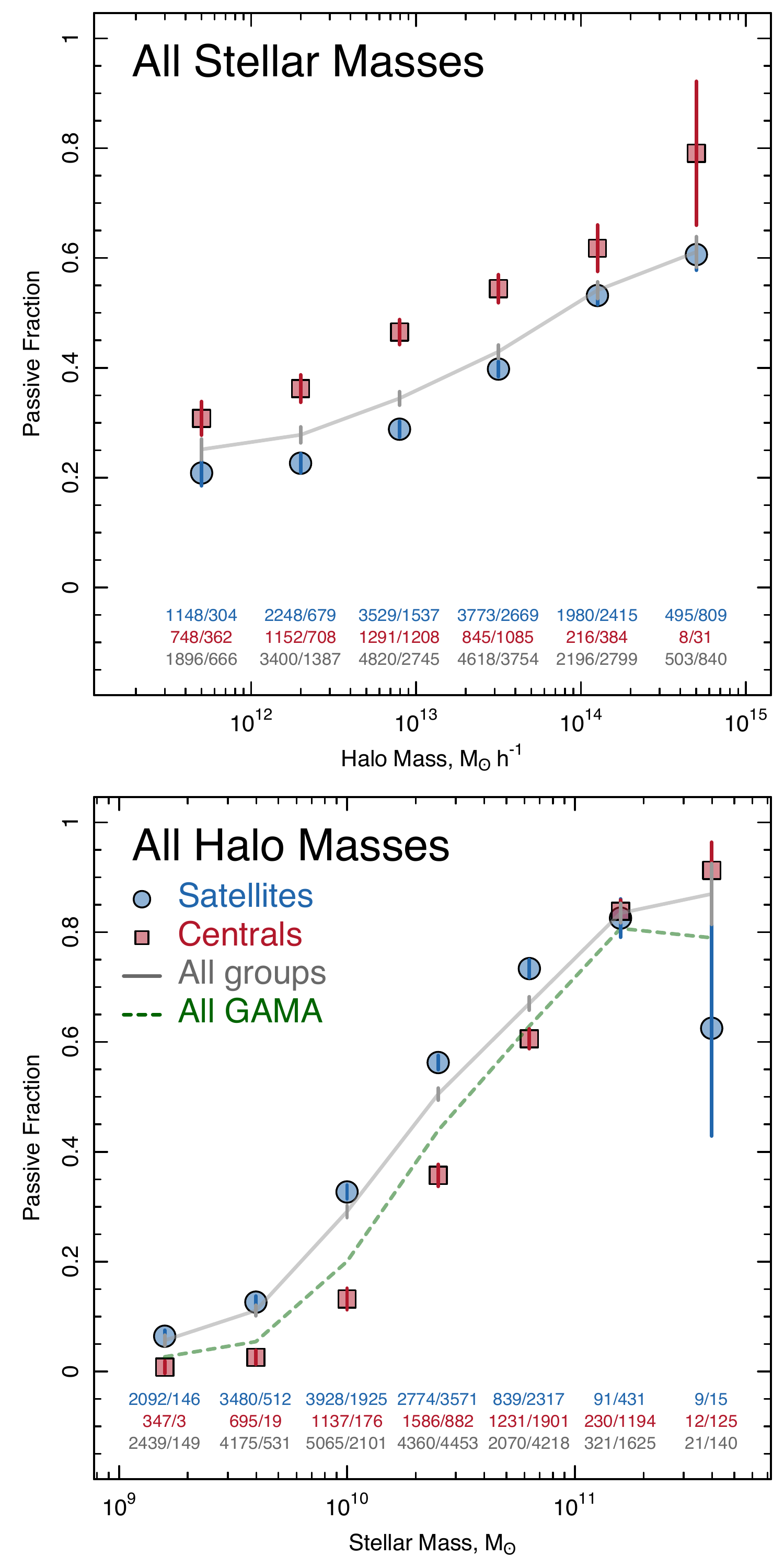}
\vspace{-4mm}

\caption{Top: The global passive fractions for all group galaxies (grey line), centrals (red) and satellites (blue) as a function of halo mass. Bottom: The global passive fractions as a function of stellar mass where we also include all GAMA galaxies irrespective of environment (green line). Error bars show the bootstrap errors plus binomial distribution errors in quadrature. Numbers show the number of star-forming/passive galaxies that go into each data point.}
\vspace{-4mm}

\label{fig:selection2}
\end{center}
\end{figure}

\subsection{Deriving Passive Fractions}
\label{sec:derive}

As with any flux-limited survey, the detectability of both galaxies and groups decreases with redshift as intrinsically faint (low-mass) galaxies can only be detected in local volumes. Within the main results presented in this paper the biases induced by this effect are minimal. This is due to the fact that we are comparing the relative quenched fraction of centrals and satellites, and these selection biases largely affect both centrals and satellites in equal measures (modulo that fact that centrals and satellites have different passive fractions and the detectability of sources varies with SFR, which is discussed later). In addition, the GAMA sample is relatively complete at $z<0.2$ to the bulk of the stellar masses studied here ($i.e.$ the sample is $>95$\% complete to log$_{10}$[M$_{*}$/M$_{\odot}$]$>$10.0 galaxies out to $z\sim0.2$). However, in order to most robustly derive the absolute passive fractions, and to be consistent with W18, we apply completeness correction weights to our sample using the standard $V_{\mathrm{t}}/V_{\mathrm{max}}$ method. 

Here we do not calculate individual weights for each galaxy, but follow the description outlined in \cite{Lange15} to avoid highly-weighted galaxies skewing the results. Firstly, we exclude all sources at log$_{10}$[M$_{*}$/M$_{\odot}$]$<$9.0, where the GAMA sample is incomplete over our redshift range and volume corrections would be large. We then split the full sample into $\Delta$log$_{10}$[M$_{*}$/M$_{\odot}$]=0.2 bins. For each bin we calculate the maximum redshift at which the lower mass end containing $>97.7$\% of the sample could be observed ($z_{\mathrm{max}}$). For each stellar mass bin we then calculate a weight using:

\begin{equation}
w(M_{\mathrm{bin}})=\frac{V_{\mathrm{t}}}{V_{z_{\mathrm{max}}}(M_{\mathrm{bin}})},
\end{equation}
        
\noindent where $w(M_{\mathrm{bin}})$ is the weight in a given bin of stellar mass, $V_t$ is the total volume of the sample, calculated as the comoving volume between $0.01<z<0.2$ and $V_{z_{\mathrm{max}}}(M_{\mathrm{bin}})$ is the comoving volume out to zmax for a given stellar-mass bin. We then set the weight of any bin where $w(M_{*})<1$ to $w(M_{*})=1$ and assign the weighting to all galaxies within each bin. This is equivalent to a volume-limited sample out to z$\sim$0.2. We note that the completeness function (and therefore $V_{z_{\mathrm{max}}}$) will in fact be different for passive and star-forming systems. However, to be consistent with W18, we use a common $V_{z_{\mathrm{max}}}$ for all galaxies within a particular stellar-mass bin.       

To calculate the passive fraction, $f_{P}$, of given subsample, $S$, we follow W18 using:

\begin{equation}
f_{P}(S)=\frac{\mathlarger{\sum^{S}_{i=1}}w_{i}(M_{\mathrm{bin}}) \times P}{\mathlarger{\sum^{S}_{i=1}}w_{i}(M_{\mathrm{bin}})},
\end{equation} 

\noindent where $w_{i}$ is the individual galaxy weight based on its stellar mass bin and $P$ is the binary passive value with $P$=1 for passive and $P$=0 for star-forming. In order to determine the effect of measurement errors for these passive fractions, we bootstrap resample for 500 iterations varying each galaxy's SFR, stellar mass and halo mass with a normal distribution using their $1\sigma$ errors taken from \cite{Davies16b} and \cite{Taylor11} for SFRs and stellar masses respectively. For halo masses, we use eq. 20 of R11 which equates group multiplicity to halo mass error.  For the errors in all of our figures we show the standard deviation of all bootstrap samples combined in quadrature with binomial distribution errors estimated using a Beta Distribution following the procedure of \cite{Cameron11}.

Initially we take galaxies at all stellar masses in our sample and compare the passive fraction of centrals and satellites as a function of halo mass, shown in the top panel of Fig. \ref{fig:selection2}. The numbers in each panel display the number of star-forming/passive galaxies that go into each data point ($i.e.$ in the lowest halo mass bin there are 2,562 sources, of which 1,452 are satellites and 1,110 are centrals, and of the satellites 1148 are star-forming and 304 are passive). We find that the passive fraction increases with halo mass for both populations and that centrals are more likely to be passive at a given halo mass, consistent with many previous results \citep[$e.g.$][and W18]{Brinchmann04, Wetzel12, Bluck16}. This is due to the fact that, at a given halo mass, a central is likely to be more massive, and hence more likely to be quenched via secular processes. In addition, the same is true for satellites (more massive haloes can host more massive satellites), but also more massive haloes are likely to have stronger satellite quenching mechanisms. Hence, multiple entangled physical effects can drive the correlations observed in this panel.     

The bottom panel of Fig. \ref{fig:selection2} then displays the halo-mass-agnostic passive fractions of centrals and satellites as a function of stellar mass ($i.e.$ over all halo masses). This shows the flipped trend, that at a given stellar mass, satellites are more likely to be quenched than centrals. This agrees with the well-known scenario of environmental quenching, highlighting that additional processes must affect satellites at a given stellar mass, increasing the relative number of passive systems \citep[$i.e.$ see][]{Weinmann09,Knobel13,Bluck16, Grootes17}. We also find that with increasing stellar mass, the relative difference in passive fraction between centrals and satellites decreases and at log$_{10}$[M$_{*}$/M$_{\odot}$]$\sim$11.0 they are the same within uncertainty. This is consistent with our current understanding, which predicts that environmental quenching is less effective at higher stellar masses, where the few massive star-forming galaxies have large gas reservoirs and can easily retain them through environmental interactions (this is discussed further in the following section). As such, the evolution of the most massive galaxies is unlikely to be strongly affected by environment ($i.e.$ central/satellite status). 

We do note that the passive fractions for satellites in our highest stellar mass bin do not follow the general trend at all other stellar masses, even when considering errors. This is unlikely to be real, which may mean that that our errors are under estimated, potentially due to our assumption of a binomial distribution. However, if we assume a Poisson distribution instead, our errors increase slightly but still show a difference between centrals and satellites at the highest stellar masses. A potential explanation for this is shot noise from small sample sizes which makes the distribution neither binomial or Poisson (there are only 24 satellites in this bin) and/or cosmic variance, as GAMA does not robustly probe a wide variety of the most massive galaxies or the most massive halos - where these satellites will reside. Therefore, we caveat that our results may not be robust for the small number of satellite galaxies at log$_{10}$[M$_{*}$/M$_{\odot}$]$\gtrsim$11.5.               

There are several results from cosmological hydrodynamic simulations that agree with the general trends in our observations. For example, \cite{Correa17} showed that the morphology of low-mass passive galaxies were varied, while those of massive galaxies were very similar (all highly bulge-dominated). The latter held true even if the massive galaxies were satellites, suggesting that the morphology of passive galaxies was not strongly affected by environment. \cite{Lagos18a} found similar trends but this time analysing the kinematics of galaxies instead of morphology. These results have been interpreted by the authors as massive satellites quenching by the same processes as massive centrals. Wright et al. (in preparation) also shows a clear dichotomy in the quenching mechanisms and timescales of satellite and central galaxies at log$_{10}$[M$_{*}$/M$_{\odot}$]$<$10.5, but similar mechanisms/timescales at higher stellar masses. All of these results have been obtained for the EAGLE simulations \citep{Crain15, Schaye15, McAlpine16}, but similar results have also been reported for Illustris-TNG \citep{Nelson18}.

\section{Central/Satellite Similarity?}

Fig. \ref{fig:selection2} may not fully encapsulate the environmental quenching processes that would be delineated by the central/satellite divide. This is due to the fact that secular quenching processes  and the passive population both vary as a function of stellar mass \citep[$e.g.$ see][]{Peng10, Weisz15, Davies15b, Davies16a, Davies18b}, with massive galaxies more likely to be passive. This is displayed as the grey (for GAMA group galaxies) and green (for all GAMA galaxies) lines in the bottom panel of Fig. \ref{fig:selection2}; $i.e.$ irrespective of central/satellite divide, more massive galaxies are more likely to be passive. In addition, the stellar mass function also varies with halo mass \citep[$e.g.$][Vazquez-Mata et al., in prep]{Yang09,Eckert16}. This can complicate crude diagnostics such as those in Fig. \ref{fig:selection2}. For example, while passive fractions of centrals are higher than satellites at all halo masses, centrals are also more likely to be massive galaxies at all halo masses. Therefore, they are more likely to be passive simply due to secular processes; irrespective of environment. As such, in order to identify potential similarity between central and satellite passive fractions (or a lack thereof) we must control for both stellar and halo mass.

\subsection{Passive fractions when controlled for stellar and halo mass}
\label{sec:controlled}

\begin{figure*}
\begin{center}
\vspace{-5mm}

\includegraphics[scale=0.58]{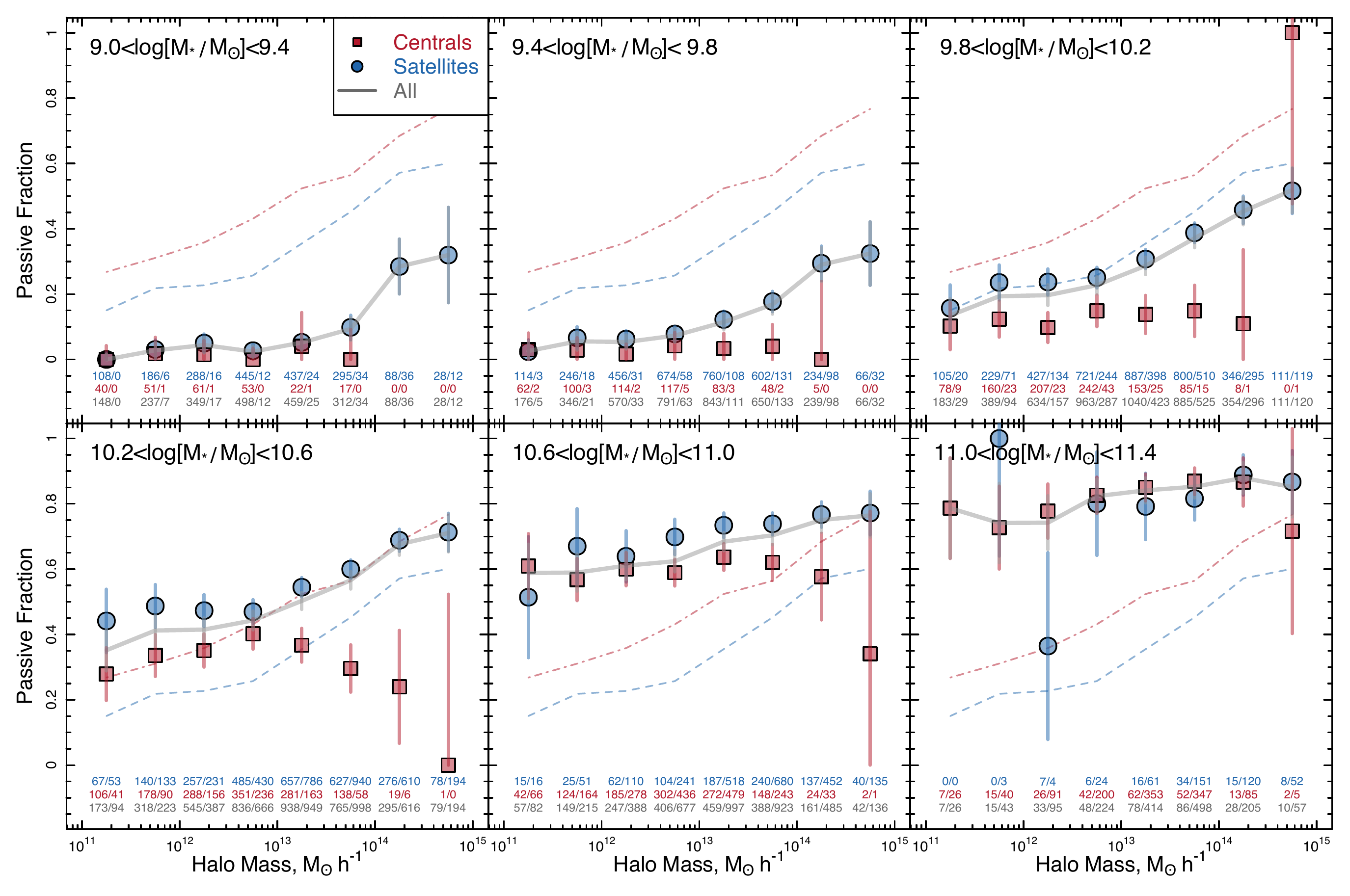}
\vspace{-2.5mm}

\includegraphics[scale=0.58]{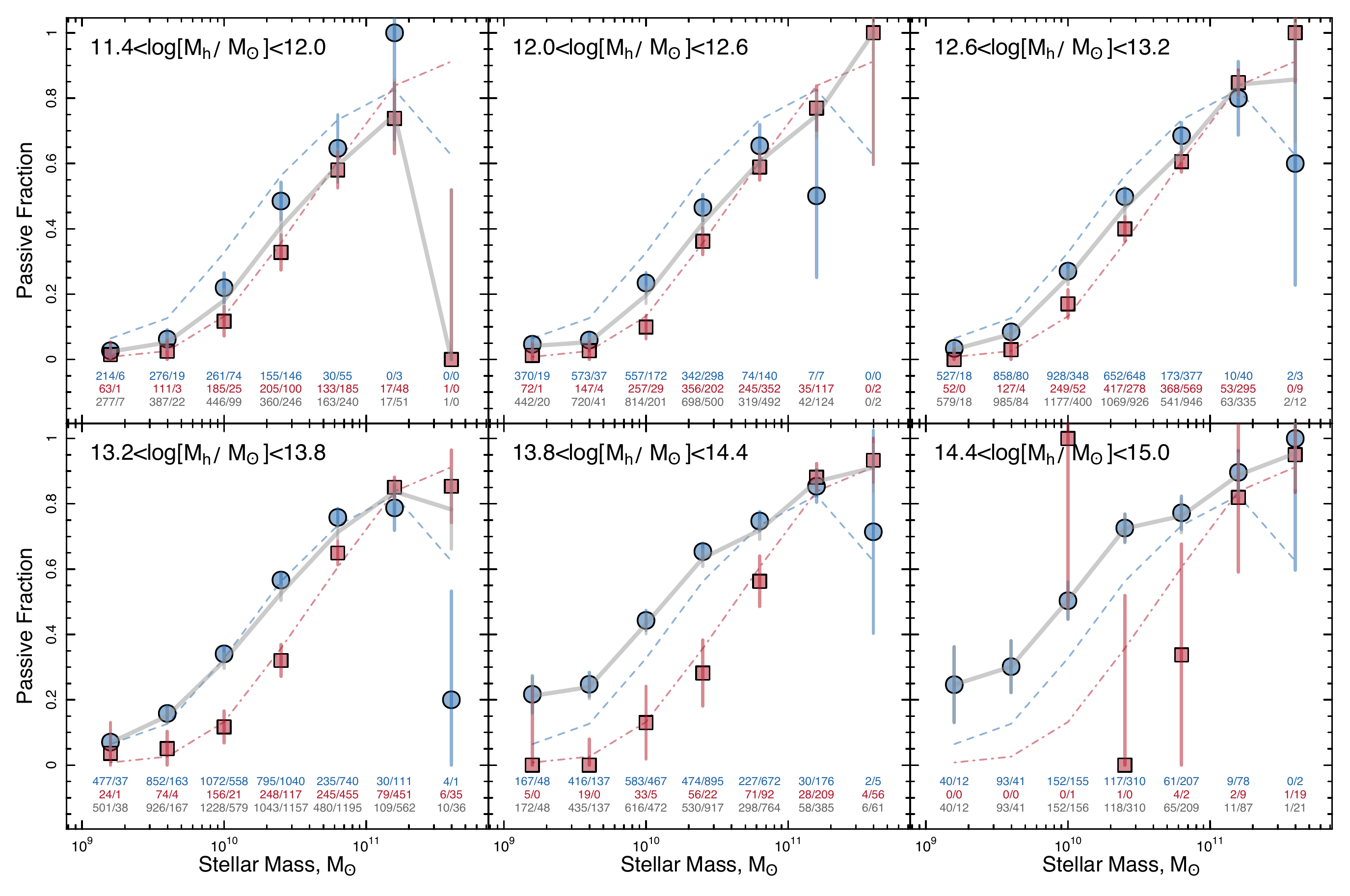}

\caption{Top panels: The passive fractions for all galaxies (grey line), centrals (red squares) and satellites (blue circles) as a function of halo mass, split into different stellar mass bins using the same ranges as W18. Bottom panels: The same as the top panels but passive fractions as a function of stellar mass, split into different halo mass bins. Error bar show the bootstrap errors plus binomial distribution errors in quadrature. Numbers at the bottom of the Fig. show the number of star-forming/passive galaxies that go into each data point. The blue dashed and red dot-dashed lines displays the global central and satellite passive fractions respectively for all stellar masses (top) and halo masses (bottom) as a reference point, taken from Fig. \ref{fig:selection2}. In all but the most massive galaxies in the most massive haloes we see a difference between the passive fractions of centrals and satellites. At a given halo mass and stellar mass (at log$_{10}$[M$_{*}$/M$_{\odot}$]$<$10.5), satellites are more likely to be passive than centrals, as expected from additional satellite quenching mechanisms. }
\label{fig:Conform}
\end{center}
\end{figure*}

We first separate our sample into six $\Delta$log$_{10}$(M$_{*}$/M$_{\odot}$)=0.4\,bins of stellar mass at 9.0$<$log$_{10}$[M$_{*}$/M$_{\odot}$]$<$11.4, consistent with W18. We then repeat the analysis described in Section \ref{sec:derive} for galaxies in each stellar mass bin. The top six panels of Fig. \ref{fig:Conform} display the passive fraction of centrals and satellites as a function of halo mass in different stellar mass bins. 

Firstly, we find that, as in the global distribution, the passive fractions for satellites increase with halo mass in all stellar mass bins except for the most massive (log$_{10}$[M$_{*}$/M$_{\odot}$]$>$11.0) galaxies. In contrast, for the majority of stellar masses, centrals appear largely agnostic to halo mass ($i.e.$ the lines are close to flat). This is interesting as it suggests that central quenching mechanisms are not strongly correlated with their larger-scale environment. A potential mechanism for central galaxy quenching is that in massive galaxies the formation of a hot corona inhibits gas outflow form stellar feedback leading to a build-up of gas in the central regions of the galaxy. This in turn triggers a response from the central black hole triggering accretion, feedback and a suppression of star-formation \citep{Bower17}. Given that the presence of hot corona is correlated with halo mass, one might expect this quenching mechanism to also be correlated with host halo. However, \cite{Bower17} also show that this affect is somewhat binary, occurring in all log$_{10}$[M$_{h}$/M$_{\odot}$]$>$12.0 haloes. As such, this mode of central quenching would be ubiquitous across almost all of the halo mass explored in our work, and we may not see any significant correlation between central passive fraction and halo mass.              

At the highest stellar masses, we find that within errors passive fractions of both centrals and satellites appear flat with halo mass. We also observe that the general trend is for a clear divide between centrals and satellites, with satellites displaying higher passive fractions. The exception to this is in the most massive galaxies in the most massive haloes (where almost all galaxies are passive) and in the lowest mass galaxies in the most lowest mass haloes (where almost all galaxies are star-forming). These results contradict the findings of W18, but are consistent with current understanding of galaxy evolution processes which suggests satellites should undergo additional quenching mechanisms, and that this quenching should be stronger in larger haloes  \citep[$e.g.$][]{vandenBosch08, Weinmann09, Wetzel12, Peng12, Knobel13,Grootes17}. We also re-observe the trend of increasing passive fractions with stellar mass, irrespective of halo mass ($i.e.$ lines move to higher massive fraction with increasing stellar mass) and that this increase is more rapid in centrals (as in Fig. \ref{fig:selection2}).

The bottom panels of Fig. \ref{fig:Conform} display the converse. We separate our sample into six $\Delta$log$_{10}$(M$_{h}$/M$_{\odot}$)=0.6\,bins of halo mass at 11.4$<$log$_{10}$[M$_{h}$/M$_{\odot}$]$<$15.0, once again with consistent binning to W18. We then show the passive fraction of centrals and satellites as a function of stellar mass for each halo mass bin. As expected, we find that at all halo masses the passive fraction increases with increasing stellar mass for both centrals and satellites. At lower halo masses (top rows), we find that both centrals and satellites have similar passive fractions. They are slightly distinct at intermediate stellar masses (10$<$log$_{10}$[M$_{*}$/M$_{\odot}$]$<$11.0), but at both lower and higher stellar masses, the passive fractions of centrals and satellites are comparable. This is consistent with the results described in the top panels of Fig. \ref{fig:Conform}, as at low stellar masses and low halo masses almost all galaxies are star-forming, and at high stellar masses and low halo masses almost all galaxies are passive.

This is also as expected, as the environmental quenching effect of these low-mass haloes is small and the passive fraction is more strongly correlated with stellar mass alone ($i.e.$ via secular processes). However, with increasing halo mass, the passive fractions of centrals and satellites at log$_{10}$[M$_{*}$/M$_{\odot}$]$<$11.0 diverge. For centrals, passive fractions remain relatively constant with increasing halo mass at close to the global value (dot-dashed line). For satellites, passive fractions increase with halo mass. Once again, this is consistent with the picture of higher mass haloes providing an increased quenching effect on satellites, but leaving centrals largely unscathed.    

Our trends are some-what similar to those observed in Fig. 3 of \cite{Hirschmann14}, which compares central/satellite passive fractions for SDSS galaxies as a function of local galaxy density (comparable to halo mass, see their Fig. 5). Our results extend to lower stellar and halo masses, and show slightly larger separation between centrals and satellites at log$_{10}$[M$_{*}$/M$_{\odot}$]$<$11.0, but over the same stellar/halo masses are comparable. By contrast, Fig. 4 of \cite{Knobel15} displays the passive fraction of centrals and satellites when matched on stellar mass (panel b), and stellar mass and local galaxy density (panel c), finding almost identical passive fractions of centrals and satellites. Their panel b can be compared to the bottom panel of Fig. \ref{fig:selection2}, which also shows central and satellite passive fractions at a given stellar mass. However, we find a separation between centrals and satellites, which is not observed by \cite{Knobel15}. In addition, their panel c can be compared to our Fig. \ref{fig:Conform}, with passive fractions controlled for stellar mass and halo mass ($i.e.$ local galaxy density). Here we also see a separation between central and satellite passive fractions at log$_{10}$[M$_{*}$/M$_{\odot}$]$<$11.0, which is not found by \cite{Knobel15}.         

While the \cite{Knobel15} results use only N$>$2 systems, and our figures display N$>$1 groups, we repeat our analysis for just N$>$2 groups and find our results do not significantly change. However, we note that \cite{Knobel15} only claim a similarity between centrals and satellites at log$_{10}$[M$_{*}$/M$_{\odot}$]$>$10.3 ($i.e.$ largely only the three highest stellar mass points in Fig. \ref{fig:selection2} and the bottom panels of Fig. \ref{fig:Conform}). This is where we also find passive fractions of centrals and satellites are similar. As such, in the overlapping stellar and halo mass ranges probed our results are not in strong contention with \cite{Knobel15} or \cite{Hirschmann14}.

\begin{figure*}
\begin{center}
\includegraphics[scale=0.6]{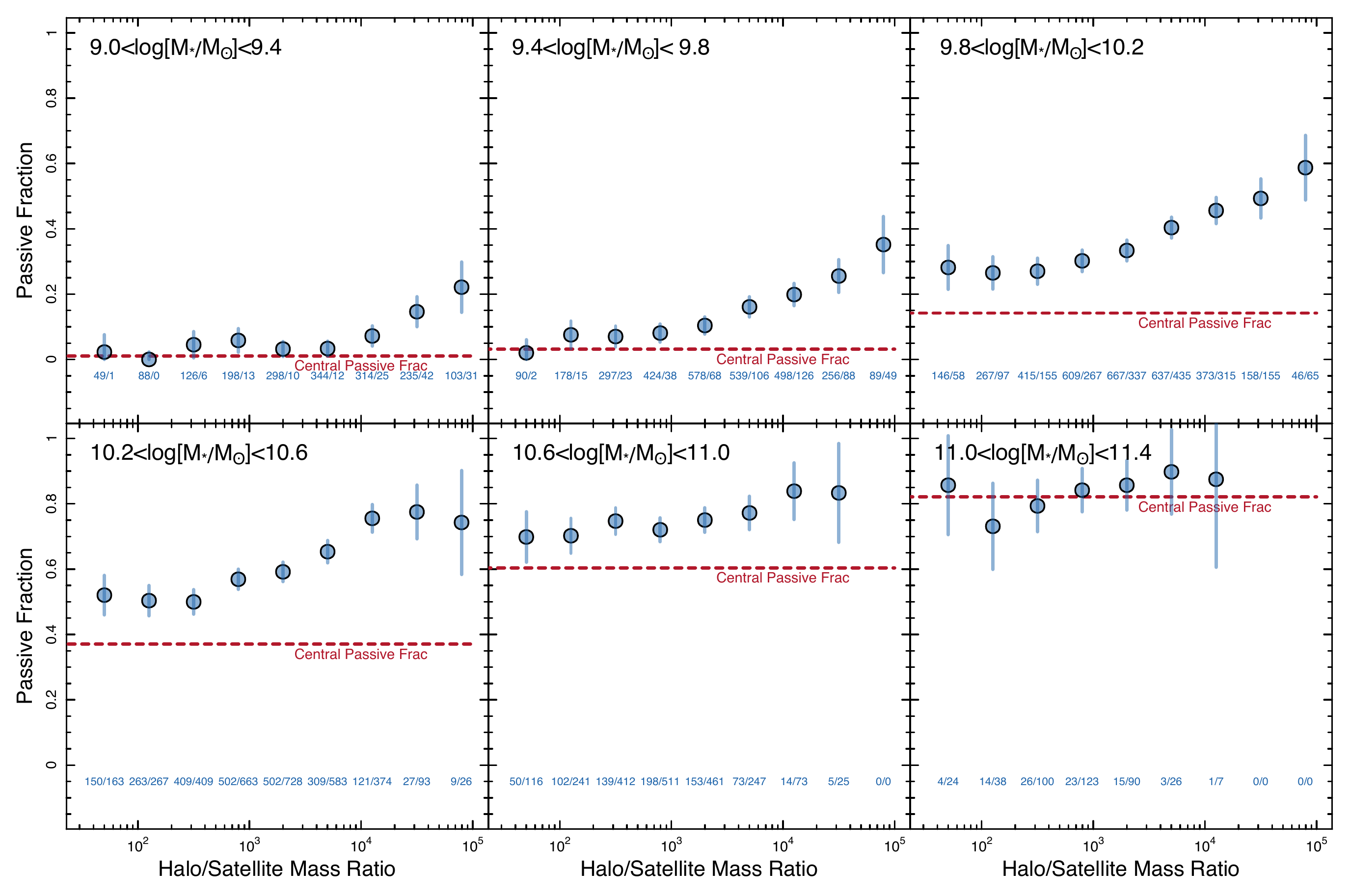}
\caption{The passive fraction of satellite galaxies as a function of total halo mass-to-satellite stellar mass ratio in $\Delta$MassRatio=0.4\,dex bins. In all panels the horizontal red dashed line displays the typical central passive fraction over the panel's stellar mass range. We see a clear trend of the passive fraction of satellites increasing with mass ratios. When a galaxy is large for its halo (halo/satellite mass ratio is low) satellites have a passive fraction that is close to centrals at the same stellar mass. These satellites could be defined as centrals in there own right and/or are largely unaffected by environmental quenching processes. As mass ratios increase, satellites become much smaller than their halo. Here environmental quenching processes, such as strangulation, stripping, harassment, and tidal interactions become progressively stronger, leading to higher passive fractions.}
\label{fig:ratio2}
\end{center}
\end{figure*}

\begin{figure*}
\begin{center}
\includegraphics[scale=0.6]{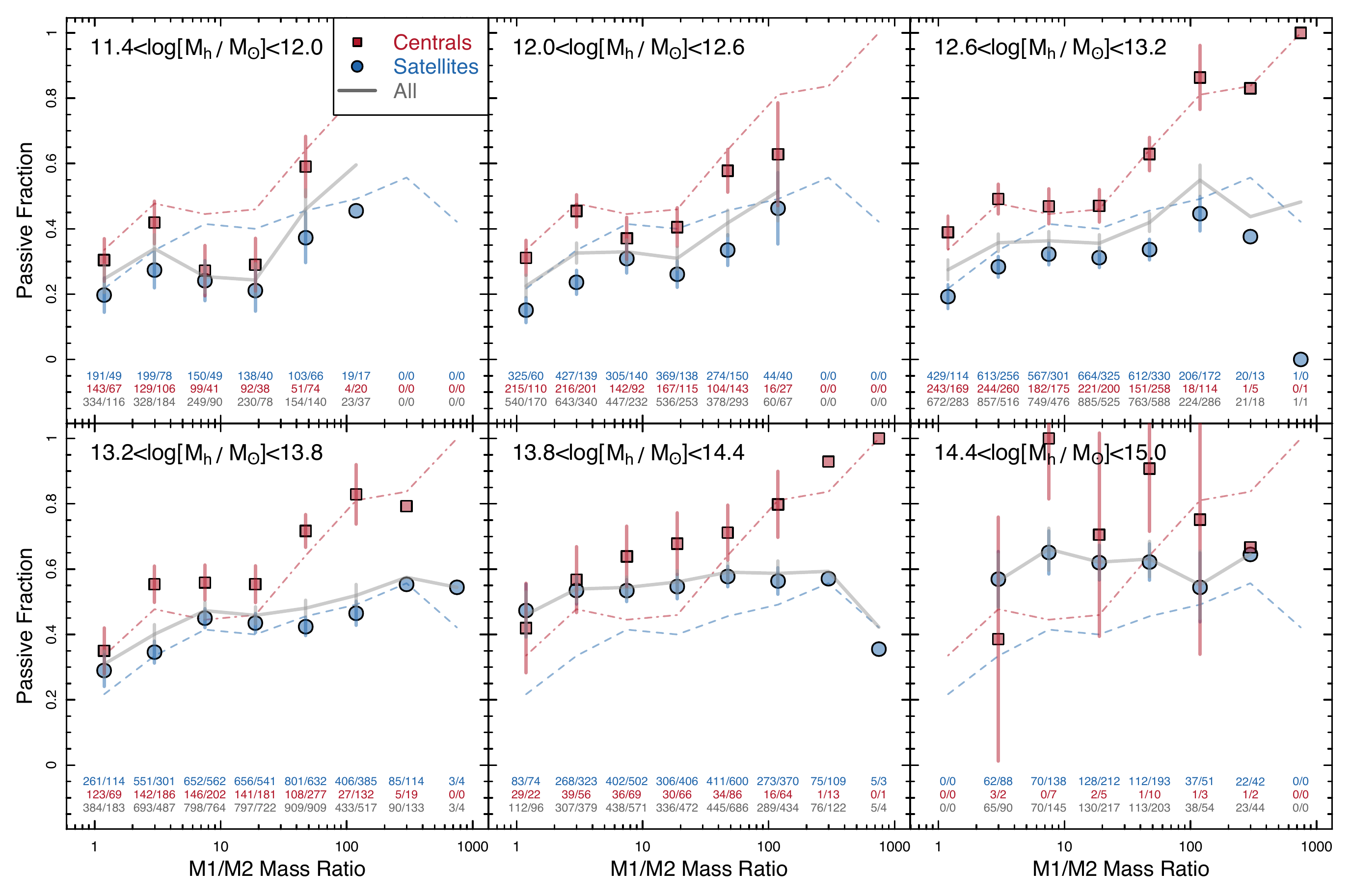}
\caption{The passive fraction of central and satellite galaxies as a function of the stellar mass ratio between the central to second-ranked galaxy (M1/M2), in a number of stellar mass ranges. This metric is a proxy for group age. The red dot-dashed and blue dashed lines show the global values for centrals and satellites respectively ($i.e.$ over all stellar/halo masses). Both centrals and satellites are more likely to be passive if they reside in older groups (larger M1/M2) and this trend is stronger for centrals than satellites. This is expected as centrals in older groups are more massive, as they have had more time to accumulate stellar mass, and older themselves (and are therefore more likely to be passive). Satellites in older groups will be a mixed bag of sources which have been in the group environment for a long time (and are therefore passive) and recently in-falling satellites (which are potentially star-forming). In contrast, young groups are more likely to only contain the recently in-falling star-forming population. This results in both centrals and satellites having increasing passive fractions with M1/M2 ratio, but the effect being more pronounced in centrals.}
\label{fig:M1M2}
\end{center}
\end{figure*}

\subsection{Passive fractions as a function of mass ratios}

Next we consider how the passive fraction of satellites varies as a function of halo mass/satellite stellar mass ratio in different stellar mass bins (Fig. \ref{fig:ratio2}). We over-plot a line displaying the central passive fraction within the given stellar mass bin. These panels highlight that at a fixed stellar mass, the passive fraction of satellites increases when the difference in mass between the satellite and halo is massive. When satellites are massive for their halo, passive fractions are close to that of centrals. This is consistent with the currently held view that environmental quenching processes are strong when a satellite is small in comparison to its host halo, as it can undergo significant strangulation, stripping, harassment, tidal processes, etc. When a galaxy is massive for its halo, environmental quenching effects will be weaker, as galaxies can retain their star-forming gas and are less affected (in terms of quenching) by any tidal interactions. This is also consistent with results from the EAGLE simulation which find that quenching timescales are shorter for higher halo/satellite mass ratios - leading to larger passive fractions (Wright et al. in preparation). 

Finally, we explore the passive fraction of centrals and satellites as a function of the mass ratio between the central and second ranked (based on stellar mass) galaxy in the group, hereafter M1/M2 (Fig. \ref{fig:M1M2}). This ratio is used as a proxy for the age of the group \citep[$e.g$][]{Ponman94, Khosroshahi04, Khosroshahi07}, as within older groups the central has consumed a larger fraction of its massive satellites. We find that, at a fixed halo mass, the passive fraction of both centrals and satellites increases with M1/M2. This suggests that galaxies, as expected, have had more time to become quenched in older groups. This is consistent with the results exploring HI gas depletion in local groups, which find that older groups are more gas poor and therefore, are more likely to host quenched populations \citep{Nichols11, Nichols13}. We also find that this trend is more pronounced for centrals than satellites, likely due to the fact that satellite populations can be constantly replenished with star-forming galaxies \citep[or rejuvenated by interaction-induced star-burst events, $e.g.$][]{Davies15b}. However, once centrals are quenched the majority may remain so ($c.f.$ central gas-rich mergers may replenish centrals with gas and induce star-formation).  

Interestingly, we also see a decrease in the passive fractions at M1/M2$\sim$20 (or equally likely an increase at M1/M2$\sim$2-3). This is observed in central and satellite galaxies both in terms of the global distribution (dashed lines) and each of the panels for log$_{10}$(M$_{h}$/M$_{\odot}$)$<$13.8, but is more pronounced in centrals. If M1/M2 correlates with with group age, this suggests that groups at a particular evolutionary stage have lower passive fraction (higher star-forming fraction) than those which are both older and younger. This is intriguing and could potentially be linked to the group relaxation timescales (although it would be unlikely to stay fixed as a function of halo mass), mergers and/or the typical galaxy quenching timescales \citep[$e.g.$ see][]{Bremer18}. Further, if we remove isolated pairs (simply $N=$2 FOF systems in the R11 group catalogue) from our sample this decrease at M1/M2$\sim$20 is still present but somewhat less pronounced, suggesting it may in part be due to galaxy-galaxy interactions. However, this observation could equally also be due to some currently unexplored selection effect. This warrants further investigation which is beyond the scope of this paper.  


In summary, our results display differences between central and satellite passive fractions over halo and stellar mass ranges where satellite quenching is likely to be an important driver of galaxy evolution. We also find that the largest differences between the central and satellite populations occur where environmental quenching is likely to be most pronounced; $i.e.$ in low-stellar-mass galaxies when controlled for halo mass, and large halo mass when controlled for stellar mass. Lastly, we show that passive fractions increase as satellites become increasingly less-massive than their halo and that both centrals and satellites are more likely to be passive in older groups. It is worth noting that our results do not take into account the location of satellite galaxies within the group, which may have a strong impact on environmental quenching mechanisms \citep[$i.e.$][]{Barsanti18}.   

To test the validity of our results we also varied the SFR indicator used, the choice of dividing line between passive and star-forming systems, and definition of central galaxy. This is described in Appendix \ref{sec:variation}. In summary,  we find that for reasonable choices the trends seen in our data remain.

\section{Discussion}

Following our results, it is interesting to consider how the previous works of W18, \cite{Hirschmann14} and \cite{Knobel15} found similar passive fractions in centrals and satellites. There are a number of possible differences between their samples and the one that we discuss here. 

Firstly, their SDSS sample does not cover the same stellar and halo mass range as our GAMA sample, and has a different completeness. The main place where this becomes apparent is in the identification of low-mass central galaxies. While centrals down to these stellar masses will be observed in SDSS, their lower mass/luminosity satellites will not. Hence, they will not be identified as an N$>$1 group. However, within GAMA our sample extends to much lower stellar masses, allowing us to identify centrals and satellites to much lower stellar masses.  Further, when considering Fig. 3 of W18, we find that there are very few centrals with high stellar to halo mass ratios at log$_{10}$[M$_{*}$/M$_{\odot}$]$<$10.2. This is where we see the largest difference between centrals and satellites in GAMA. In fact, if we only consider our data points in bins which contain points in Fig. 3 of W18, we would largely see similarity between the centrals and satellites. This is also found to be true for both \cite{Hirschmann14} and \cite{Knobel15}, as discussed in Section \ref{sec:controlled}.  

Secondly, the W18 work uses a relatively strict selection of passive systems at 1\,dex below the SFS. Considering the top left panel of their Fig. 1, we can see that this selection may include some of the passive cloud as star-forming galaxies. In Appendix \ref{sec:variation}, we show that by decreasing the selection between passive and star-forming systems, we can artificially produce similar passive fractions between centrals and satellites. \cite{Hirschmann14} use a simple sSFR$<10^{11}$\,yr$^{-1}$ cut, which assumes that sSFR-M$_{*}$ relation is flat (which it is not, and therefore can bias the division of  star-forming and passive systems as function of stellar mass), while \cite{Knobel15} identify the tough point in density between the star-forming and passive populations by eye. This results in a selection that is very similar to ours.             

Lastly, the all of these previous studies use groups identified from the halo-based method of Y07, while within GAMA we use the friends-of-friends algorithm of R11. A number of recent studies have found that many results based on environment are very susceptible to the method of group finding \citep{Campbell15}. For example, \cite{Kafle16} found that the potential stellar mass segregation observed in SDSS using the Y07 groups could not be reproduced in GAMA for the G$^{3}$C, irrespective of choice of central. \citet{Kafle16} attribute this to subtle differences in the group finders which may have an affect here. To parametrise this, they apply both the Y07 and R11 group finders to the EAGLE simulation and find that the R11 method more accurately reproduces the intrinsic EAGLE groups. They note that, as discussed in \cite{Duarte15}, it is potentially the computation of luminosity incompleteness during the Y07 group finding which propagates to the abundance matching technique. This then leads to the incorrect estimate of group masses.

\begin{figure*}
\begin{center}
\includegraphics[scale=0.65]{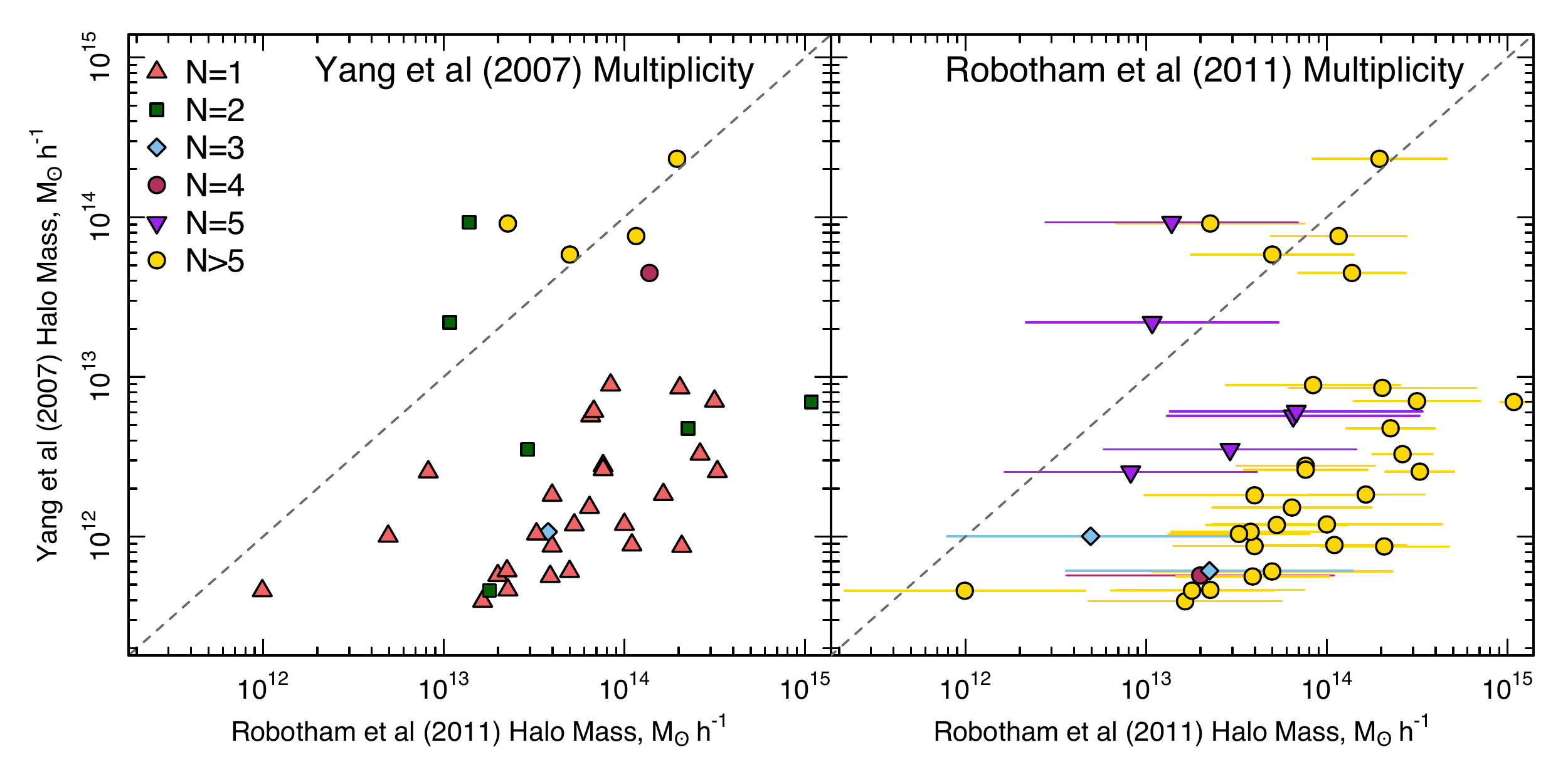}
\caption{Comparison of the halo masses estimated from the Y07 and R11 group finders for a common set of groups. The left panel displays points colour coded by the group multiplicity of the Y07 catalogue (given in the legend) and the right panel colour coded by the group multiplicity of the R11 catalogue. Errors for the R11 are estimated using N and the dashed grey line shows a 1:1 correlation. Where both catalogues contain N$>$3 members ($i.e.$ at M$_{h}$(Y07)$\gtrsim10^{13.5}$\,M$_{\odot}$\,h$^{-1}$) the halo masses agree. However, at lower masses in the Y07 catalogue there is an offset between the halo masses. In this regime the  Y07 groups typically contain just $N=$1 member and thus have their halo mass estimated from the stellar mass-halo mass relation, while R11 groups have N$>$4 members and therefor are likely to have more robust halo mass estimates. }
\label{fig:Yang}
\end{center}
\end{figure*}

To partially explore this further, we compare the halo mass estimates from Y07 and R11 in a common subsample of groups that are identified in the same volume. As GAMA covers a sub-area of SDSS a number of groups appear in both catalogues. To perform this matching we identify all Y07 groups where the physical group centre falls within the co-moving extent of a GAMA group (as defined by the radius contains all group members, R100 in R11). Using this conservative selection, we identify just 39 groups. Fig. \ref{fig:Yang} displays the halo masses of these common groups from both group finders. The left panel is colour coded by Y07 group multiplicity, while the the right panel is coloured by  R11 multiplicity. In the regime where groups in both catalogues contain N$>$3 members ($i.e.$ at M$_{h}$(Y07)$\gtrsim10^{13.5}$\,M$_{\odot}$\,h$^{-1}$) the halo masses agree well. However, at lower halo masses in the Y07 catalogue there is a systematic offset between the group finders. For these groups the Y07 catalogue typically contains just one member. Given their abundance matching procedure, the halo mass from such systems will essentially be assigned based on the theoretical central stellar mass-to-halo mass relation. However, within the R11 catalogue, the same groups have N$>$4 members and therefore, are likely to have a robust more measurement of their halo mass based on the group satellite velocity dispersion. In addition, we also note that the R11 halo masses at M$_{h}\gtrsim10^{13.0}$\,M$_{\odot}$\,h$^{-1}$ ($i.e.$ almost all of the groups in Fig. \ref{fig:Yang}) have been independently verified via the weak lensing analysis of the  Kilo-Degree Survey (KiDS) team \citep{Viola15}. This difference in halo masses has important consequences for the results presented in the previous works exploring similarity between passive fractions of centrals and satellites as a function of halo mass. Potentially the centrals (and some satellites) of M$_{h}\lesssim10^{13.5}$\,M$_{\odot}$\,h$^{-1}$ groups in these analyses are, in reality, residing in M$_{h}\gtrsim10^{13.5}$\,M$_{\odot}$\,h$^{-1}$ groups.                     

\vspace{1mm}

As noted previously, one of the other main differences between these group finders is that the Y07 method assigns isolated centrals to haloes based on abundance matching. To partially explore this, we assign a halo mass to all isolated centrals in GAMA using the analytic form of the central stellar mass to halo mass relation taken from \cite{Behroozi13}, and then repeat our analysis (see Appendix \ref{sec:singles}). We find that, while the number of central galaxies dramatically increases (from 10,141 N$>$1 group centrals to 52,669 isolated and group centrals), the overall central passive fractions do not change significantly (see Fig. \ref{fig:singles}). This is due to the fact that these isolated centrals are predominantly at the low stellar- and halo-mass end of our samples (where passive fractions are already close to zero), and are almost exclusively star-forming. Therefore, they do little to decrease the passive fraction. This potentially suggests that the potential differences between our results and those of W18, \cite{Hirschmann14} and \cite{Knobel15} are not due to the additional isolated centrals in the Y07 catalogue.        

Interestingly, in a follow up paper to W18, \citet{Wang18b} investigate the passive fractions of centrals and satellites in both the L-GALAXIES model and EAGLE simulations. They find little (EAGLE) or no (L-GALAXIES) intrinsic similarity between central and satellite passive fractions. However, when they apply the Y07 group finder to the simulations, they increase the similarity. As such, it is likely that at least a part of their result is driven by theses subtleties of the group finding, which we do not see here. An interesting test for the W18 result, would be to repeat their analysis with different SDSS-based group catalogues, such as that of \cite{Saulder16}. \cite{Kafle16} found that the lack of mass segregation observed in GAMA was similar to those obtained from the SDSS group catalogue of \cite{Saulder16}, but different to those of Y07, again highlighting the significant impact the group-finding process can have on results.

\section{Conclusions}

We have derived the passive fractions for central and satellite galaxies in the GAMA G$^{3}$C group catalogue, as a function of halo and stellar mass. We find that:\\
\\

\noindent $\bullet$ When considering all stellar masses (top panel Fig. \ref{fig:selection2}), passive fractions of both centrals and satellites increase with halo mass. This is as expected because higher mass haloes contain larger/older galaxies which are more likely to be passive. We also find that passive fractions are higher for centrals than satellites at a given halo mass, as centrals are likely to be more massive at a given halo mass.  \\       
\\

\noindent $\bullet$ When considering all halo masses (bottom panel Fig. \ref{fig:selection2}), passive fractions of both centrals and satellites increase with stellar mass. This is also as expected because higher mass galaxies are more likely to be passive. Here, passive fractions are higher for satellites than centrals at a given stellar mass (for log$_{10}$[M$_{*}$/M$_{\odot}$]$<$11.0 galaxies). This supports previous well-documented evidence for environmental quenching of low-mass satellites. \\
\\

\noindent $\bullet$ When controlled for both stellar mass and halo mass (Fig. \ref{fig:Conform}) we still find a difference between centrals and satellite passive fractions. This disagrees with some of the conclusions from the SDSS work of W18, \cite{Hirschmann14} and \cite{Knobel15}. We also find that this quenching is more pronounced in satellites of lower stellar mass and in haloes of higher mass.\\
\\

\noindent $\bullet$ We find that satellite passive fractions increase with halo/satellite mass ratio (Fig. \ref{fig:ratio2}), consistent with the picture of environmental quenching processes such as tidal stripping and harassment being stronger when mass ratios are large.   \\
\\

\noindent $\bullet$ We show that passive fractions in both centrals and satellites increase with M1/M2 ratio (Fig. \ref{fig:M1M2}), which is a proxy for the age of the group. This increase is more pronounced in centrals than satellites. We attribute this to the fact that, once quenched, centrals predominantly remain so, whereas satellite populations can be replenished with star-forming systems.\\
\\

\noindent $\bullet$ We explore how varying our SFR indicator, separation between the star-forming and passive population and choice of central affect the central/satellite passive fractions. We find that for reasonable choices the trends seen in our data remain (Appendix \ref{sec:variation}).\\
\\    

\noindent $\bullet$ Finally we suggest that the similarity between central and satellites observed in previous studies is likely due to subtleties in the Y07 group finding (most likely halo mass estimates), but may be contributed to by both the stellar/halo mass range probed by SDSS, and/or choice of star-forming/passive selection. However, we also find that when including isolated centrals, as in the Y07 catalogue, the difference between centrals and satellites remains.  \\
\\    

Our results form a consistent picture of satellite quenching in group environments which is in agreement with many pervious studies \citep[$e.g.$][]{vandenBosch08, Weinmann09, Wetzel12, Peng12, Knobel13,Grootes17}. In this picture, once controlled for stellar mass, halo mass has weak-to-no correlation with the star-forming properties of centrals (in M$_{h}\gtrsim10^{11.4}$\,M$_{\odot}$ haloes), $i.e.$ passive fractions of centrals are mostly flat with halo mass (Fig. \ref{fig:Conform}). Simulations, such as those used in \cite{Gabor10}, suggest that to produce the observed red sequence and luminosity function, this quenching of centrals must occur either via feedback from intense star-formation and AGN following major mergers \citep[$e.g.$][]{Springel05, Cox06, Hopkins06}, mostly occurring at high redshift, or the triggering of AGN feedback via a hot corona \citep[as discussed previously, $e.g.$][]{Bower17}. Neither of these process are unlikely to be aligned with halo mass at z$\sim0$, but may be correlated with total stellar mass ($i.e.$ to first order, more massive galaxies have had more mergers).       

In contrast, when controlled for stellar mass, satellite galaxies are strongly impacted by their environment, suggesting a different evolutionary path to centrals \citep[$e.g.$][]{Wetzel12, Hahn17}. This impact is more pronounced in high mass haloes (Fig. \ref{fig:Conform}) and when satellites are small in comparison to their host halo (Fig. \ref{fig:ratio2}). These are the regimes where environmental quenching processes such as, tidal and ram-pressure stripping \citep{Gunn72, Moore99, Poggianti17,Brown17, Barsanti18} and/or harassment \citep{Moore96} are likely to be stronger. Centrals, which sit at the centre of their halo and are almost exclusively the largest galaxy in the group, suffer far less from these processes.            

Finally, both centrals and satellites are more likely to be passive in older groups (Fig. \ref{fig:M1M2}), and this effect is more pronounced for centrals then satellites. For centrals, this is likely to be simply due to formation age, $i.e.$ centrals in older groups are themselves older and have had more time to consume their star-forming gas. While the same is true for satellites, the satellite population in old groups is also replenished with younger star-forming galaxies, leading to a flattening of this relation.

\vspace{1mm}

In summary, combining these results our analysis is consistent with a model where the star-forming properties of all galaxies are correlated with stellar mass, with more massive/older galaxies having both consumed more of their star-forming gas, and more likely having taken part in major-merger quenching event. In centrals, this is the dominant quenching mode, and as the galaxies reside in the centre of their haloes, they are not strongly affected by environmental processes. This manifests as a strong correlation between central passive fraction and stellar mass, but little correlation with halo mass. In satellites, additional environmental quenching mechanisms ($i.e.$ tidal and ram-pressure stripping and/or harassment) affect their star-formation properties. These quenching processes are likely to be most efficient in the most over-dense environments and when a satellite is low-mass in comparison to it's host halo. This is seen as a correlation between passive fraction and halo mass, and passive fraction and halo/satellite mass ratio.        

\section*{Acknowledgements}

GAMA is a joint European-Australasian project based around a spectroscopic campaign using the Anglo- Australian Telescope. The GAMA input catalogue is based on data taken from the Sloan Digital Sky Survey and the UKIRT Infrared Deep Sky Survey. Complementary imaging of the GAMA regions is being obtained by a number of in- dependent survey programs including GALEX MIS, VST KiDS, VISTA VIKING, WISE, Herschel-ATLAS, GMRT and ASKAP providing UV to radio coverage. GAMA is funded by the STFC (UK), the ARC (Australia), the AAO, and the participating institutions. The GAMA website is \url{http://www.gama-survey.org/}.

CL has received funding from a Discovery Early Career Researcher Award (DE150100618) and by the ARC Centre of
Excellence for All Sky Astrophysics in 3 Dimensions (ASTRO 3D), through project number CE170100013.

\appendix

\section{Variation with SFR indicator, SF/Passive selection, and Central Identification}
\label{sec:variation}

\begin{figure*}
\begin{center}
\includegraphics[scale=0.6]{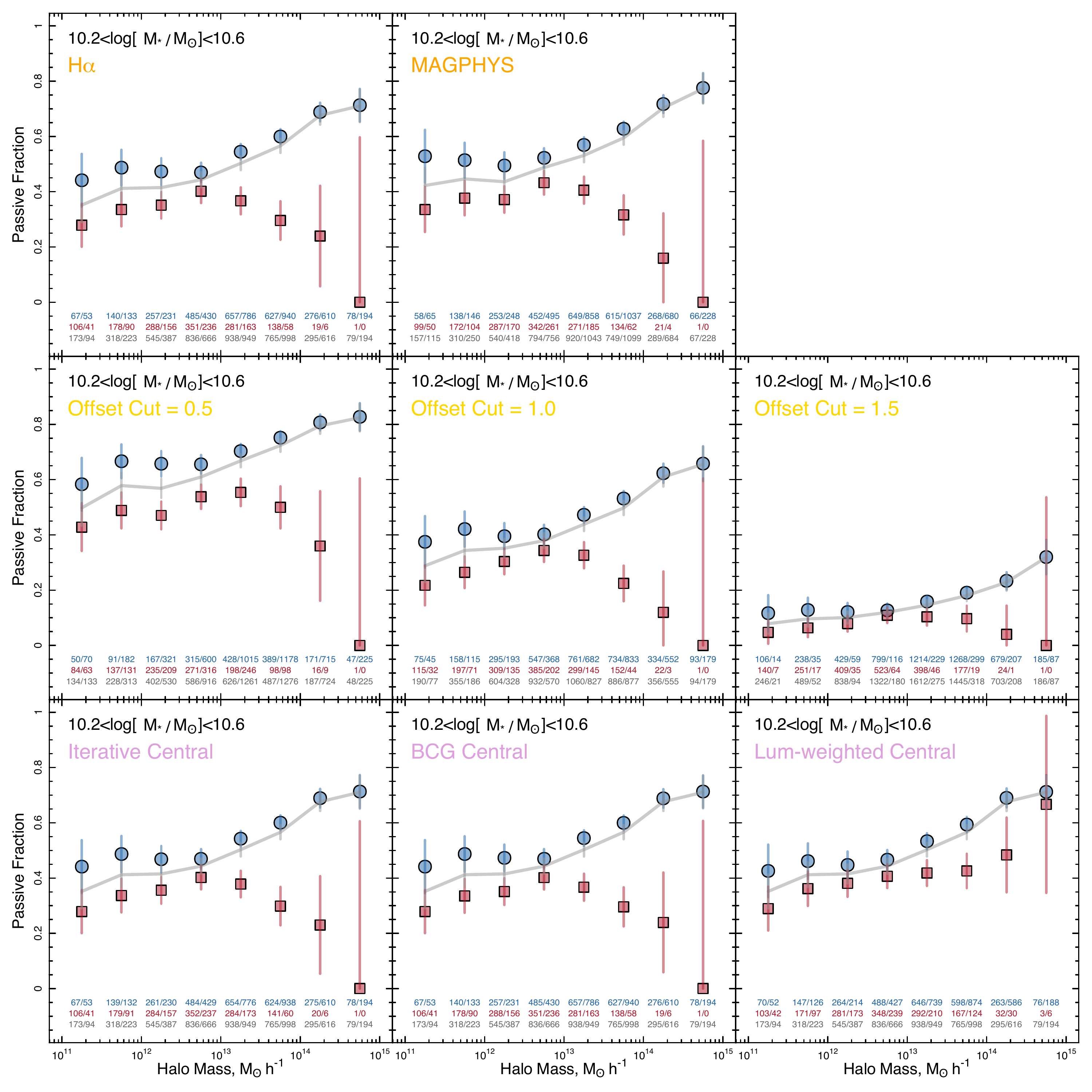}
\caption{How varying our method affects the central/satellite passive fractions as a function of halo mass for 10.2$<$log$_{10}$[M$_{*}$/M$_{\odot}$]$<$10.6 galaxies. Top row shows differences between H$\alpha$ and \textsc{magphys} SFRs (with iterative central and SF/Passive divide explained in the text). Middle row shows changes based on the dividing line between passive and star-forming systems, where Offset Cut is the offset from the SFS in dex (with iterative central and H$\alpha$ SFRs). Bottom row shows changes based on choice of central (with H$\alpha$ SFRs and SF/Passive divide explained in the text).}
\label{fig:comp}
\end{center}
\end{figure*}

In order to explore the validity of our results, we repeat our analysis using a different SFR indicator, vary our star-forming/passive selection, and use each of the central definitions from the GAMA group catalogue. Fig. \ref{fig:comp} displays a number of examples of how varying these parameters affects the central/satellite passive fractions. Here we only show the passive fractions as a function of halo mass for 10.2$<$log$_{10}$[M$_{*}$/M$_{\odot}$]$<$10.6 galaxies ($i.e.$ the bottom left panel of the halo mass plots in Fig. \ref{fig:Conform}).   

Firstly, using SED-derived SFRs from \textsc{magphys} instead of H$\alpha$ we find that our results do not significantly change and all trends seen in the data remain (top row Fig. \ref{fig:comp}). Thus it is unlikely that our choice of SFR indicator is significantly driving our results.

Next we vary the selection boundary between passive and star-forming systems. For this we return to a selection using the linear offset from the SFS as in W18, and apply offsets between 1.5\,dex and 0.5\,dex as our dividing line. When using relatively small offsets ($<1.0$\,dex) we still do not see similarity between centrals and satellites. At larger offsets we do begin to see some similarity between centrals and satellites (middle row Fig. \ref{fig:comp}). However, this is due to the fact that we are just including a significant fraction of the passive cloud in the star-forming sample (which is clearly incorrect). 

Finally, we also use the different methods for central identification from the G$^{3}$C catalogue described in Section \ref{sec:Groups} (bottom row Fig. \ref{fig:comp}). In the majority of cases the iterative group central is the same as the BCG \citep[which is used in the rest of this paper, see][]{Robotham11} and therefore the results using this metric do not change significantly from our previous results. However, when using the luminosity-weighted centre we do see the passive fraction becoming similar between centrals and satellites. R11 argue that the luminosity weighted centre provides a poor estimation of the group central galaxy as it simply takes the galaxy closest to the centre of light. If the galaxy selected as the group central in this manner is in fact a satellite, we will not expect to see any differences between centrals and satellites. In addition, the luminosity weighting in the G$^{3}$C catalogue is performed in the $r$-band, and therefore may weight more heavily to star-forming than passive systems, artificially biasing the passive fractions.

In summary, with reasonable assumptions for SFRs, selection between passive and star-forming galaxies, and choice of group central, we still see a discrepancy between centrals and satellites when controlled for stellar and halo mass, consistent with our current understanding of galaxy evolution process in group environments.            

\section{Inclusion of isolated centrals}
\label{sec:singles}

As discussed in this paper, one of the main differences between the Y07 and R11 group finders is that the Y07 work assigns halo masses to isolated centrals based on abundance matching. To test whether this is the driving factor in the W18 results, we repeat our analysis when including isolated centrals. As the R11 catalogue does not contain halo masses for isolated galaxies, we use the functional form of the central stellar mass-to-halo mass relation from \cite{Behroozi13}. We then assign all isolated centrals a halo mass based on their stellar mass and include a $\pm$0.25\,dex random error. We note that this is a relative crude approach, but will provide some clarification as the effect isolated centrals on our results.  

\begin{figure*}
\begin{center}
\includegraphics[scale=0.6]{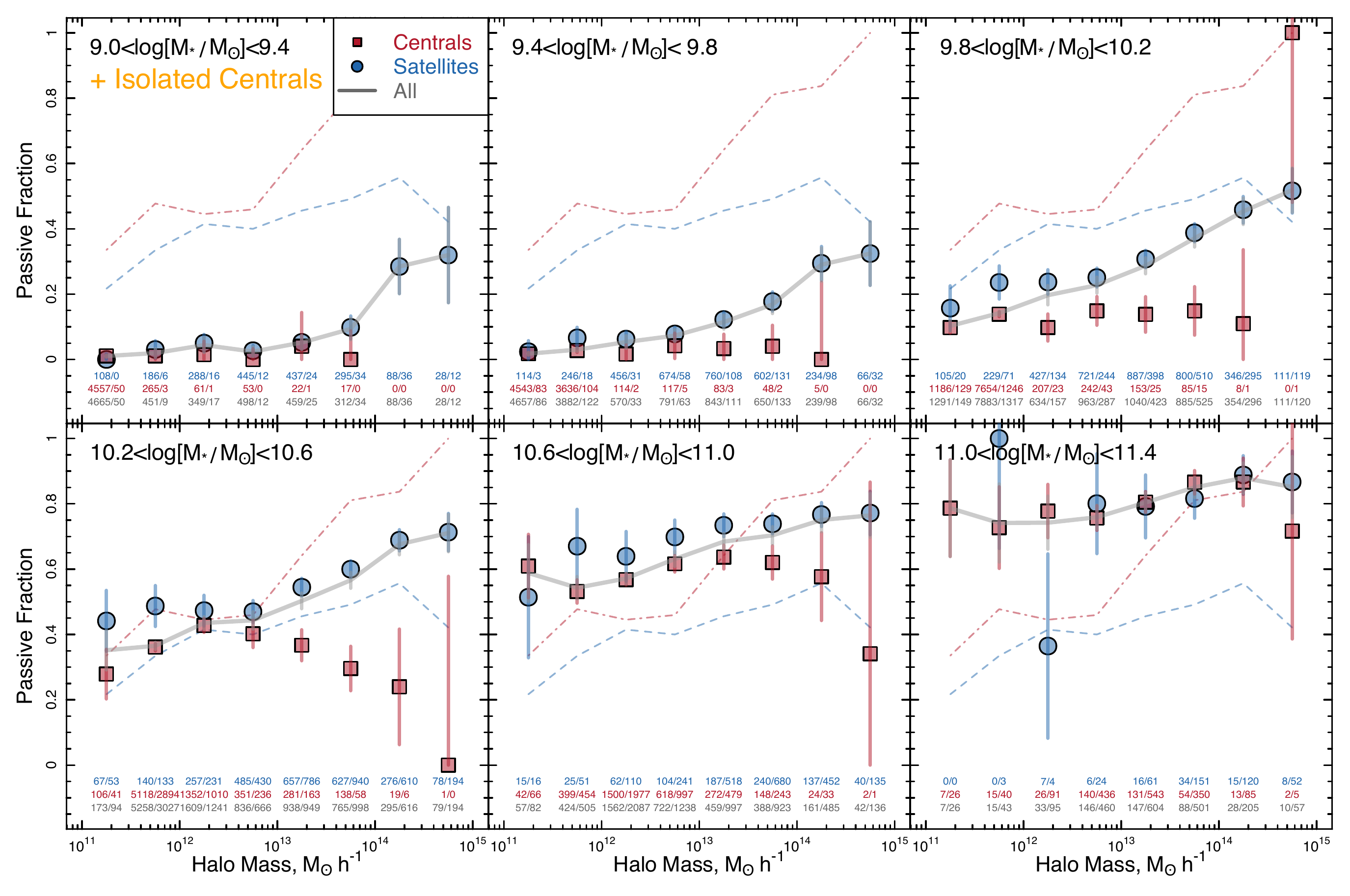}
\caption{The same as the top panels of Fig. \ref{fig:Conform} but also including isolated centrals. Halo masses for these centrals are defined using the analytic form of the stellar mass-halo mass relation outlined in \citet{Behroozi13}. While the number of centrals dramatically increases in our sample, the passive fractions change very little, potentially suggesting that the W18 results are not driven by the inclusion of isolated centrals in the Y07 catalogues.}
\label{fig:singles}
\end{center}
\end{figure*}

Fig. \ref{fig:singles} displays the same as the top panels of Fig. \ref{fig:Conform}, but including these centrals. While the total number of central galaxies increase dramatically from 10,141 centrals in N$>$1 systems to 52,669 N$>$0 centrals, the central passive fractions change very little. This is largely due to the fact that the majority of isolated centrals are at low stellar masses and are star-forming. At this point passive fractions are very close to zero, and therefore can not be changed by the addition of more star-forming galaxies. This potentially highlights that the difference between our results and those of W18 are not due to the inclusion of isolated centrals in the Yang et al catalogue.            

\bsp	
\label{lastpage}
\end{document}